\documentclass[amsmath,amssymb,twocolumn,aps,nofootinbib]{revtex4-1}

\usepackage{amssymb}
\usepackage{multirow}
\usepackage{amsfonts}
\usepackage{amsmath}
\usepackage{amssymb}
\usepackage{graphicx}
\usepackage{dcolumn}
\usepackage{times}
\usepackage{color}
\usepackage{epsf}

\newcommand{\nin}{\noindent}
\newcommand{\be}{\begin{equation}}
\newcommand{\ee}{\end{equation}}
\newcommand{\bea}{\begin{eqnarray}}
\newcommand{\eea}{\end{eqnarray}}

\newcommand{\nn}{\nonumber\\}

\newcommand{\clA}{\mathcal{A}}
\newcommand{\dmu}{\partial_\mu}
\newcommand{\dnu}{\partial_\nu}

\newcommand{\dunu}{\partial^\nu}
\newcommand{\kk}{\Bbbk}
\newcommand{\vb}{\vec{\beta}}

\begin{document}

\title{Gauss' Law and Non-Linear Plane Waves for Yang-Mills Theory}

\author{A. Tsapalis$^{a,b}$, E.P. Politis$^a$, X.N. Maintas$^a$ and F.K. Diakonos$^a$}


\affiliation{
{$^a$ Department of Physics, University of Athens, GR-15771 Athens, Greece}\\
{$^b$ Hellenic Naval Academy, Hatzikyriakou Avenue, Pireaus 185 39, Greece}
}

\date{\today}

\begin{abstract}
We investigate Non-Linear Plane-Wave solutions of the classical Minkowskian Yang-Mills (YM) equations of motion. By imposing a suitable ansatz which solves Gauss' law for the $SU(3)$ theory, we derive solutions which consist of Jacobi elliptic functions depending on an enumerable set of elliptic modulus values. The solutions represent periodic anharmonic plane waves which possess arbitrary non-zero mass and are exact extrema of the non-linear YM action. Among them, a unique harmonic plane wave with a non-trivial pattern in phase, spin and color is identified. Similar solutions are present in the $SU(4)$ case while are absent from the $SU(2)$ theory.

\end{abstract}
\maketitle

\section{Introduction}

Classical solutions of field theories remain of central interest in the understanding of the structure and dynamics of the Standard Model particle interactions.
Regarding the pure Yang-Mills (YM) theory, major attention has been drawn on the extrema of the Euclidean action --the instantons-- whose structure and properties are determined by topology (for a Review see~\cite{books}). A lot of work has been produced on the relevance of the instanton configurations to the confining properties of the YM quantum vacuum as well as the quark dynamics coupled to the gauge bosons. On the other hand, in the Minkowskian 3+1-D spacetime, solitary waves, solitons or travelling localized lumps with a finite energy are forbidden in the $SU(N)$ gauge theory as Coleman has shown generally~\cite{Coleman1} 
that finite-energy non-singular gauge field configurations that do not radiate out to spatial infinity are not allowed to exist. Minkowskian Non-Linear Plane Waves (NLPW) have been shown to exist for the $SU(2)$ theory~\cite{Savvidy1} and have the form of the Jacobian Elliptic function ${\rm cn}(\omega \,t- \vec{k} \cdot \vec{x} ;1/2)$ with fixed elliptic parameter $1/2$ and relativistic dispersion relation $ \omega^2 =\vec{k}^2 +m^2 $ for an arbitrary mass parameter $m$. The self-interaction term is reflected in the anharmonic form of the ${\rm cn}$-wave while scale invariance leaves free the value of mass $m$. Other solutions of the $SU(2)$ Minkowskian equations of motion (EoM) are generated from the massless scalar $\phi^4$ theory extrema via the Corrigan-Fairlie-'t Hooft-Wilczek ansatz~\cite{Corrigan},\cite{Ohteh1} or generalized forms~\cite{Ohteh2}.

Regarding the $SU(3)$ theory, a restrictive form of massless plane waves were initially shown to exist by Coleman~\cite{Coleman2}, and in a more general form in~\cite{melia}, technically dropping out the interaction. A class of massive plane-wave solutions of the non-abelian theory can be constructed from classical
solutions of the massless $\phi^4$ theory~\cite{Frasca}. These are anharmonic waves of the ${\rm cn}(\omega \,t- \vec{k} \cdot \vec{x} ;1/2)$ type and are 
essentially the waves contructed via the 'multiple copies trick' 
~\cite{Smilga}. Such solutions have also been shown to become relevant to the properties of the quantum theory in the strong coupling limit~\cite{Frasca}.

It is the purpose of this work to investigate if more general NLPW solutions exist for the $SU(3)$ gauge theory EoM. For this reason we present in Section 2 
the detailed form of the EoM for the general $SU(N)$
NLPW ansatz based on Lorentz symmetry and the scale invariance of the theory. In Section 3 we review the so-called 'multiple copies trick' ~\cite{Smilga} which determines $SU(N)$ solutions proportional to the ${\rm cn}(\omega \,t- \vec{k} \cdot \vec{x} ;1/2)$ field with appropriate Non-Abelian constant factors and derive in particular the form of the constants for the $SU(3)$ theory.
In Section 4 we propose a more general ansatz that solves the Gauss law constraint equations for the $SU(3)$ gauge theory. We arrive to a set of coupled cubic 
equations for complex fields which correspond to planar point particle dynamics  bounded by the $r^4$ potential. The gauge field color and polarization indices of the solution are mixed in a non-trivial scheme.
The general solution is fixed by the angular momentum $L$ of the particle in the sense that the elliptic parameter $k^2$ of the Jacobian Elliptic functions is connected to $L$ via a non-linear equation. It is interesting that for the highest
value of $L$ allowed, a harmonic massive plane wave is shown to exist solving the 
$SU(3)$ EoM even in the presence of the interaction terms. In Section 5 we show that plane waves other than the ones in~\cite{Savvidy1} do not exist for the $SU(2)$ theory. We finally outline the relevant ansatz for $SU(4)$ and the embedding of the $SU(3)$ solutions in it. The possible utility of such solutions is commented in the final section.

\section{Non linear plane waves and the Yang-Mills equations of motion}

\nin
The Langrangian density of the YM theory is defined via
\be
\mathcal{L}=-\frac{1}{4}\mathcal{F}_{\mu \nu}^{a} \mathcal{F}^{\mu \nu a}
\label{eqA1}
\ee
where $\mathcal{F}_{\mu \nu}^a$ is the antisymmetric field tensor of the gauge field $\mathcal{A}_{\mu}^a$ ($\mu, \nu = 0,1,2,3$ are spacetime indices with a $(1,-1,-1,-1)$ metric assumed everywhere) :  
\be
\mathcal{F}_{\mu \nu}^{a}=\dmu {\clA_{\nu}^a}-\dnu{\clA_{\mu}^a + g\; f_{a b c} \clA_{\mu}^b \clA_{\nu}^c}
\label{eqA2}
\ee
The structure constants $f_{a b c}$ define the Non-Abelian $SU(N)$ algebra via the commutators of the $N^2-1$ generators $T^a$ in the fundamental $(N \times N)$ representation:
\be
[ T^a, T^b]= i\;f_{abc} T^c 
\label{eqA3}
\ee
The corresponding classical EoM for the gauge field 
are~\footnote{In covariant form $\mathcal{D}_\mu \mathcal{F}^{\mu \nu} = 0$, with the covariant derivative defined as $\mathcal{D}_\mu = \dmu - i\;g \clA_\mu$ and 
 $-i g \mathcal{F}_{\mu \nu}= [\mathcal{D}_{\mu} ,\mathcal{D}_{\nu} ]$}:
\be
\dmu \mathcal{F}^{\mu \nu a} +g  f_{a b c} \; \clA_{\mu}^b \mathcal{F}^{\mu \nu c}=0
\label{eqA4}
\ee
which in component form read explicitly:
\bea
&&\Box \clA^{\nu a}-\dunu \dmu \clA^{\mu a} +\nn && g\,f_{a b c} \Big[ (\dmu \clA^{\mu b}) \clA^{\nu c} + 2 \clA^{\mu b} \dmu \clA^{\nu c} -\clA^{\mu b} \dunu \clA_{\mu}^{c})   \Big]+\nn &&g^2\; f_{a b c} f_{c d e} \, \clA_{\mu}^b \clA^{\mu d} \clA^{\nu e} = 0 \; .
\label{eqA5}
\eea
The $\nu = 0 $ set of the above equations constitute the Gauss law constraint obeyed by the non-dynamical fields $A_0^a$, while the $\nu =1,2,3$ eqs provide the evolution of the dynamical spatial components.
We introduce generic plane wave solutions, i.e. fields depending explicitly on the
plane wave phase 
\be
\xi =\omega \,t- \vec{k} \cdot \vec{x} 
\label{eqA6}
\ee
with the momentum four-vector $\Bbbk^{\mu} = (\omega, \vec{k})$ satisfying the dispersion relation $ \omega^2 =\vec{k}^2 +m^2 $ for an arbitrary mass parameter $m$.
In the physical system of units, the gauge field also has the dimension of mass so we
scale the fields as follows, eliminating at the same time completely the coupling $g$ from the classical EoM:
\be
\clA_{\mu}^a = \frac{m}{g} \, A_{\mu}^a (\xi) \, .
\label{eqA7}
\ee
Space-time derivatives are replaced by derivatives with respect to $\xi$ (denoted by dotting):
\be
\dmu A_{\nu}^a = \kk_{\mu} \dot{A}_{\nu}^{a}(\xi) \, ,
\label{eqA8}
\ee
and the equations~(\ref{eqA5}) become:
\bea
&&m^2 \ddot{A}^{\nu a} - \kk^\nu \kk_{\mu} \ddot{A}^{\mu a} + \nn
&& m f_{a b c} \Big[\kk_{\mu} \dot{A}^{\mu b} A^{\nu c} + 2 \kk_{\mu} A^{\mu b} \dot{A}^{\nu c}  - \kk_{\nu} A^{\mu b} \dot{A}_{\mu}^{c}\Big]+ \nn
&& m^2 \; f_{a b c} f_{c d e} \, A_{\mu}^b A^{\mu d} A^{\nu e} = 0 \; .
\label{eqA9}
\eea
Now we make use of Lorentz covariance of eq.~(\ref{eqA9}) by boosting the vector fields $A_{\mu}^a$ on the proper time frame, where $\kk_{\mu} \rightarrow (m, \vec{0})$ and $\xi \rightarrow m t$, via a Lorentz transformation $\Lambda(\vb)$ with boost parameters
\be
\vb = \frac{\vec{k}}{\omega} \hspace{5mm}, \hspace{5mm} \gamma = \frac{\omega}{m} 
\label{eqA10}
\ee
and explicit spin-1 representation
\be
\Lambda^{\mu}_{\nu}(\vb)=\begin{pmatrix}
\gamma & -\vb \gamma \\
-\vb \gamma & \delta_{ij}+ \frac{\gamma^2}{\gamma+1}\beta_i \beta_j \;.
\end{pmatrix}
\label{eqA11}
\ee
The Gauss law equation becomes in the proper frame~\footnote{For convenience we use the same symbol for the rotated fields $\Lambda^{\mu}_{\nu} A^{\nu a}\rightarrow A^{\mu a}$.}:
\be
f_{a b c} A_j^b \dot{A}_j^c - f_{a b c} f_{c d e} A_j^b A_j^d  A^{0 e} = 0
\label{eqA12}
\ee
while the dynamical equations read (Latin indices $i,j =1,2,3$):
\bea
&&\ddot{A}^{a}_i + f_{a b c}\big[\dot{A}^{0 b} A^{c}_i + 2 A^{0 b} \dot{A}^{c}_i\big] + \nn && f_{a b c} f_{c d e} \big[A^{0 b} A^{0 d} -A_j^b A_j^d \big] A^{e}_i = 0 \;.
\label{eqA13}
\eea
Next, we use the remaining $t$-dependent gauge freedom to fix $A^a_0 = 0$.
The gauge group element $g(t)$ which solves the equation 
\be
g A_0^a T^a g^{-1}- i\frac{dg}{dt}g^{-1}=0
\label{eqA14}
\ee 
is formally provided by the Polyakov line, 
$g(t)=\mathcal{P}e^{-i \int^t A_o^a T^a} $ and the equations become:
\bea
&& f_{a b c} A_j^b \dot{A}_j^c = 0 \; , \hspace{4mm} (\rm{Gauss\;\; law})   \nn
&& \ddot{A}^{a}_i -f_{a b c} f_{c d e} A_j^b A_j^d A^{e}_i = 0
\label{eqA15}
\eea
Introducing also the matrices ${\bf A_i} = A^a_i T^a$ and the matrix-vector potential $ \vec{{\bf A}} = A^a_i T^a \hat{e}_i$ 
($\hat{e}_1, \hat{e}_2, \hat{e}_3$ an $R^3$ basis)  eqs.~(\ref{eqA15}) are written as
\be
\hspace{-1.2cm}
[\vec{{\bf A}} , \dot{\vec{{\bf A}}}] = 0 
\label{eqA16}
\ee
\be
\ddot{\bf A}_i + \sum_{j \neq i}[{\bf A_j},[{\bf A_j},{\bf A_i}]]= 0
\label{eqA17}
\ee
The above set of equations maintain only the global $SU(N)$ rotations,
$\vec{{\bf A}} \rightarrow g \vec{{\bf A}} g^{-1}$. The chromoelectric and chromomagnetic fields for the above configurations are easily obtained:
\be
E^a_i = \frac{m^2}{g}\dot{A}^a_i \hspace{5mm}, \hspace{5mm} 
B^a_i = \frac{m^2}{2g} \epsilon_{i j k} f_{a b c} A^b_j A^c_k 
\label{eqA18}
\ee

\section{The multiple copies technique}

\nin
A standard trick which solves the Gauss law constraint (for any $SU(N)$) is the
so-called {\it multiple copies} technique~\cite{Smilga}. 
This is the selection of copies
of a color-independent field $(\Phi_x, \Phi_y, \Phi_z)$ as
\be
(A^a_x, A^a_y, A^a_z) = (C^a_x\Phi_x, C^a_y\Phi_y, C^a_z \Phi_z) 
\label{eqB1}
\ee
for some constant vectors $C^a_x, C^a_y, C^a_z$.
Due to the antisymmetric structure of $f_{a b c}$, each of the three terms in Gauss law
\be
f_{a b c} \Big[ C^b_xC^c_x \Phi_x \dot{\Phi}_x +C^b_yC^c_y \Phi_y \dot{\Phi}_y 
+C^b_z C^c_z \Phi_z \dot{\Phi}_z \Big] = 0
\label{eqB2}
\ee
vanishes independently.
Of course the dynamical equations have to be consistent for all color indices
$a$ and these impose restrictive algebraic constraints on the $3 (N^2-1)$ constants  $C^a_i$.   
Even in this case, the equations 
\bea
&&C^a_x \ddot{\Phi}_x - f_{a b c} f_{c d e} C^e_x\Big[C^b_yC^d_y \Phi_y^2 +  
C^b_zC^d_z \Phi_z^2 \Big] \Phi_x =0 \nn
&&C^a_y \ddot{\Phi}_y - f_{a b c} f_{c d e} C^e_y\Big[C^b_xC^d_x \Phi_x^2 +  
C^b_zC^d_z \Phi_z^2 \Big] \Phi_y =0 \nn
&&C^a_z \ddot{\Phi}_z - f_{a b c} f_{c d e} C^e_z\Big[C^b_xC^d_x \Phi_x^2 +  
C^b_y C^d_y \Phi_y^2 \Big] \Phi_z =0 
\label{eqB3}
\eea
will in general present chaotic behaviour~\cite{Savvidy2}. Integrability is expected only for 
the diagonal case $\Phi_x = \Phi_y = \Phi_z= \Phi$ in which case
the compatibility of the system
\bea
&&C^a_x \ddot{\Phi} - f_{a b c} f_{c d e} C^e_x\Big[C^b_yC^d_y +  
C^b_zC^d_z \Big] \Phi^3 =0 \nn
&&C^a_y \ddot{\Phi} - f_{a b c} f_{c d e} C^e_y\Big[C^b_xC^d_x +  
C^b_zC^d_z \Big] \Phi^3 =0 \nn
&&C^a_z \ddot{\Phi} - f_{a b c} f_{c d e} C^e_z\Big[C^b_xC^d_x +  
C^b_y C^d_y \Big] \Phi^3 =0 
\label{eqB4}
\eea
still allows a large space of constants $C^a_i$ that can be traced
numerically. We investigated in particular the $SU(3)$ group and based 
on insight from section IV we confirmed that the following (not unique) 
structure
\bea
&&C^a_x = \big( {\rm cos}\phi_1,{\rm sin}\phi_1,0,0,0,0,0,0 \big) \nn
&&C^a_y = \big( 0,0,0,{\rm cos}\phi_2,{\rm sin}\phi_2,0,0,0 \big) \nn
&&C^a_z = \big( 0,0,0,0,0,{\rm cos}\phi_3,{\rm sin}\phi_3,0 \big)
\label{eqB5}
\eea
with arbitrary constant angles $\phi_1,\phi_2, \phi_3$ satisfies~(\ref{eqB4}) and leads to a single equation for $\Phi$:
\be
\ddot{\Phi} + \frac{1}{4} \Phi^3 = 0
\label{eqB6}
\ee
which is solved by 
\be
\Phi(\xi)= {\rm sn}[\frac{1}{2}\xi;-1]= {\rm cn}[\frac{1}{2}\xi;\frac{1}{2}]
\label{eqB7}
\ee
For the $SU(2)$ theory, the choise $C^1_x = C^2_y = C^3_z = 1$ and all others zero, leads to the original solution presented in~\cite{Savvidy1} with $A^1_x = A^2_y=A^3_z = \Phi$ and $\Phi$ the solution of 
\be
\ddot{\Phi} + 2 \Phi^3 = 0 \;.
\label{eqB8}
\ee

\section{SU(3)}

\nin
We present here a more general way to solve the Gauss law constraint (eq.~\ref{eqA16}) for the $SU(3)$ theory. The idea is to arrange the matrix-vector $\vec{{\bf A}}$ in such a way that it becomes orthogonal
to the matrix-vector multiplication with itself. This can be achieved by
``staggering'' the color fields of the fundamental $ 3 \times 3$ matrix along
the three orthogonal vectors of the $R^3$ basis ($\hat{e}_1, \hat{e}_2, \hat{e}_3$) in the following way:
\be
\vec{\bf A} = \begin{pmatrix}
D_3 \hat{e}_3         & \Psi_{1}^* \hat{e}_1  & \Psi_{2}  \hat{e}_2 \\ 
\Psi_{1} \hat{e}_1    & D_2 \hat{e}_2           & \Psi_{3}^* \hat{e}_3 \\
\Psi_{2}^* \hat{e}_2  &\Psi_{3} \hat{e}_3      & D_1 \hat{e}_1 
\end{pmatrix}
-\frac{1}{3}\Big( D_1\hat{e}_1+ D_2\hat{e}_2 +D_3 \hat{e}_3\Big) {\bf{I}}_3
\label{eqC1}   
\ee
with $D_1,D_2,D_3$ real functions and $\Psi_1, \Psi_2, \Psi_3$ complex functions of $\xi$. Essentially, each complex $SU(3)$ root pair --which corresponds to a doublet of non-diagonal color fields-- is aligned along one of the three spatial directions.
The explicit connection with the octet fields is given below (with all other components zero):
\bea
&&A^1_x + i A^2_x = 2\Psi_1 \nn
&&A^4_y - i A^5_y = 2\Psi_2 \nn
&&A^6_z + i A^7_z = 2\Psi_3 \nn 
&&A^3_x= 0, \,\, A^3_y = -D_2 , \,\, A^3_z = D_3 \nn
&&A^8_x= -\frac{2D_1}{\sqrt{3}}, \,\,  A^8_y= \frac{D_2}{\sqrt{3}},\,\,   A^8_z= \frac{D_3}{\sqrt{3}}
\label{eqC2}   
\eea
A direct check of the Gauss law is straightforward, since the trace piece (second term in~(\ref{eqC1})) drops out of the commutator. By construction each column (row) of the first term in~(\ref{eqC1})) is orthogonal to any other column (row) and thus the Gauss law is written:
\be
G^a T^a = \vec{\bf A}\cdot \dot{\vec{{\bf A}}} - \dot{\vec{{\bf A}}} \cdot \vec{{\bf A}} =
2 i  \begin{pmatrix}
L_1-L_2 & 0 & 0 \\
0 & L_3-L_1 & 0 \\
0 & 0 & L_2 - L_3 \end{pmatrix}
\label{eqC3}   
\ee
where we introduced the real quantities
\bea
L_1 = \frac{i}{2} (\Psi_1 \dot{\Psi_1^*} - \Psi_1^* \dot{\Psi_1}) \nn 
L_2 = \frac{i}{2} (\Psi_2 \dot{\Psi_2^*} - \Psi_2^* \dot{\Psi_2}) \nn
L_3 = \frac{i}{2} (\Psi_3 \dot{\Psi_3^*} - \Psi_3^* \dot{\Psi_3}) 
\label{eqC4}
\eea 
Implementation of the Gauss law requires:
\be
L_1 = L_2 = L_3 = L
\label{eqC5}   
\ee
with the classes of solutions distinguished from now on by the $L\ne 0$ and $L=0$ cases.

\nin
The dynamical EoM (24 in total) are derived according to~(\ref{eqC1}). We separate them in to three different groups.\\
Group 1:
\bea
\ddot{\Psi_1}+\Psi_1 \Big[ |\Psi_2|^2 + |\Psi_3|^2 + D_2^2 + D_3^2  \Big] = 0 \nn
\ddot{\Psi_2}+\Psi_2 \Big[ |\Psi_1|^2 + |\Psi_3|^2 + D_1^2 + D_3^2  \Big] = 0 \nn
\ddot{\Psi_3}+\Psi_3 \Big[ |\Psi_1|^2 + |\Psi_2|^2 + D_1^2 + D_2^2  \Big] = 0 
\label{eqC6}   
\eea
Group 2:
\bea
\ddot{D_1} + 6 D_1 |\Psi_2|^2 = 0  \;\;,\;\; \ddot{D_1} + 6 D_1 |\Psi_3|^2 = 0 \nn
\ddot{D_2} + 6 D_2 |\Psi_1|^2 = 0  \;\;,\;\; \ddot{D_2} + 6 D_2 |\Psi_3|^2 = 0 \nn
\ddot{D_3} + 6 D_3 |\Psi_1|^2 = 0  \;\;,\;\; \ddot{D_3} + 6 D_3 |\Psi_2|^2 = 0 
\label{eqC7}   
\eea
Group 3:
\bea
D_1 \Psi_1 \Psi_2 = 0   \;\;\;, \;\;\; D_1 \Psi_1 \Psi_3 = 0 \nn
D_2 \Psi_1 \Psi_2 = 0   \;\;\;, \;\;\; D_2 \Psi_2 \Psi_3 = 0 \nn
D_3 \Psi_1 \Psi_3 = 0   \;\;\;, \;\;\; D_3 \Psi_2 \Psi_3 = 0 
\label{eqC8}   
\eea

\subsection{$L\ne 0$ solutions}

\nin
Admitting $L\ne 0$ requires $\Psi_1 \ne 0, \Psi_2 \ne 0$ and $\Psi_3 \ne 0$.
From the 'Group 3' equations we are lead to 
\be
D_1 = D_2 = D_3 = 0 \;\;,
\label{eqD1}
\ee
and the coupled system of equations:
\bea
\ddot{\Psi_1}+\Psi_1 \Big[ |\Psi_2|^2 + |\Psi_3|^2 \Big] = 0 \nn
\ddot{\Psi_2}+\Psi_2 \Big[ |\Psi_1|^2 + |\Psi_3|^2 \Big] = 0 \nn
\ddot{\Psi_3}+\Psi_3 \Big[ |\Psi_1|^2 + |\Psi_2|^2 \Big] = 0 
\label{eqD2}
\eea
The above system has the interpretation of three coupled planar point dynamics
on the $\Psi_1, \Psi_2$ and $\Psi_3$ planes respectively (which are the $1-2,\; 4-5$ and $6-7$ planes in the adjoint representation). 
One may use
equivalently a polar coordinate description 
\bea
\Psi_1 = r_1 e^{i \theta_1} \;\;, \;\;\Psi_2 = r_2 e^{i \theta_2} \;\;, \;\;\Psi_3 = r_3 e^{i \theta_3} \;
\label{eqD3}
\eea
with  $(r_i, \theta_i)$ functions of the phase $\xi$.
The dynamics on each plane is invariant under independent global $U(1)$ rotations
\be
\Psi_1 \rightarrow \Psi_1 e^{i \omega_1} \;\;, \Psi_2 \rightarrow \Psi_2 e^{i \omega_2} \;\;, \Psi_3 \rightarrow \Psi_3 e^{i \omega_3} \;\;, 
\label{eqD4}
\ee
or equivalently under independent $SO(2)$ rotations on 
the $(r_i, \theta_i)$ planes.
From this we identify $L_1,L_2,L_3$ in eq.~(\ref{eqC3}) as the {\it conserved angular momenta} of the three coupled rotating particles,
\bea
L_1 = A^1_x \dot{A^2_x}-A^2_x \dot{A^1_x} = r_1^2 \dot{\theta_1} \nn 
L_2 = A^5_y \dot{A^4_y}-A^4_y \dot{A^5_y} = r_2^2 \dot{\theta_2} \nn 
L_3 = A^6_z \dot{A^7_z}-A^7_z \dot{A^6_z} = r_3^2 \dot{\theta_3} 
\label{eqD5}
\eea 
Imposing the Gauss law $L_1 = L_2 = L_3$, relates the complex functions 
$\Psi_1, \Psi_2, \Psi_3$ via linear transformations. These are well known,
e.g.
\be
\Psi_2 = a \Psi_1 + b \Psi_1^* \;\;, \;\;\; aa^*-bb^* = 1
\label{eqD6}
\ee
where $a, b$ arbitrary complex constants. Equivalently, in the cartesian 
form they act as $SL(2,R)$ transformations~\footnote{The explicit relation is $
a = \frac{1}{2} (\alpha + \delta - i (\beta-\gamma)) \;,
b = \frac{1}{2} (\alpha - \delta + i (\beta+\gamma))$. }, e.g.
\be
\Psi_2 = \begin{pmatrix} A^4_y \\ -A^5_y \end{pmatrix} = 
\begin{pmatrix} \alpha & \beta  \\ \gamma & \delta \end{pmatrix} \cdot
\begin{pmatrix} A^1_x \\ A^2_x \end{pmatrix} = \Pi \; \Psi_1
\label{eqD7}
\ee
where ${\rm det}(\Pi) = \alpha \delta - \beta \gamma = 1$.

\begin{figure}[htbp]
\centerline{\includegraphics[width=\columnwidth]{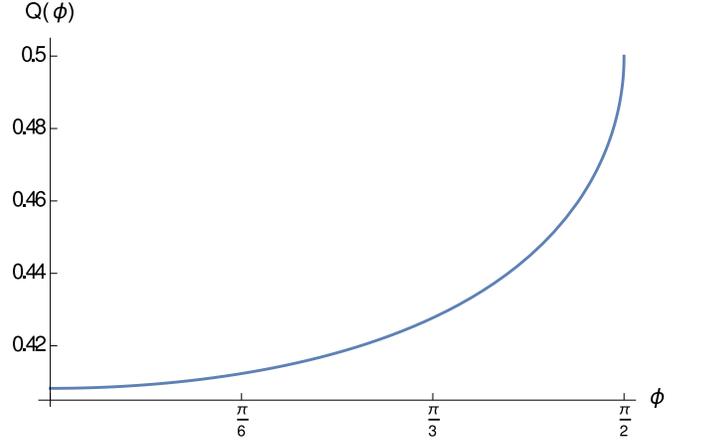}}
\caption{The function ${\cal{Q}}(\phi)$, Eq.(\ref{eqD19}), as a function of $\phi$. Each rational value of ${\cal{Q}}$ provides a solution of the $SU(3)$ EoM.}
\label{fig:fig1}
\end{figure}

The linear relations in conjuction with the dynamical equations~(\ref{eqD2})
constrain further the amplitudes:
\be
|\Psi_1|^2 = |\Psi_2|^2 = |\Psi_3|^2 = r^2\;\;.
\label{eqD8}
\ee
Up to constant angles, this also leads to $\theta_1 = \theta_2 =\theta_3=\theta$ and  
thus, we conclude that the $L\ne 0$ class of solutions is described by a 
single planar rotor bound by a central potential $V(r)= r^4/2$ and 
satisfying the EoM:
\be
\ddot{r}-\frac{L^2}{r^3}+2 r^3 =0 \;\;.
\label{eqD9}
\ee
The system is integrable via the energy constant
\be
E = \frac{1}{2} \dot{r}^2 + \frac{L^2}{2 r^2} + \frac{1}{2}r^4 \;.
\label{eqD10}
\ee
Defining the rescaled function $u$ via
\be
r^2 = \sqrt{\frac{8E}{3}} u\big((\frac{8E}{3})^{1/4} \xi\big)
\label{eqD11}
\ee
it satisfies
\be
\dot{u}^2 = 3 u - 4 u^3 - \lambda  \;\;, \;\;\;\lambda = 4 L^2 \;.
\label{eqD12}
\ee
We scaled the constant $(8 E/3)^{1/4} = 1$ in~(\ref{eqD11}) by absorbing it from now on in the mass parameter $m$. 
Eq.~(\ref{eqD12}) is solved by a {\it{Weierstrass elliptic function}}, $\cal{P}$, which is a doubly periodic function on the complex plane. We
look for real, positive, bounded solutions to describe periodic closed orbits
on the $(r, \theta)$ plane and the suitable solution is expressed in terms of the Jacobi elliptic 
functions ${\rm sn}(\xi;k^2), {\rm cn}(\xi;k^2), {\rm dn}(\xi;k^2)$ of elliptic modulus $k$ (or elliptic parameter $k^2$),
\be
u(\xi) = e_1 + (e_2-e_1) \frac{1}{{\rm dn}^2(\alpha \xi;k^2)} \;\;, 
k=\sqrt{\frac{e_3-e_2}{e_3-e_1}} 
\label{eqD13}
\ee 
The parameter $\alpha=\sqrt{e_3-e_1}$ and 
$e_1, e_2, e_3$ are the three real roots of $3u - 4u^3 -\lambda = 0.$ 
Three real roots exist only for $ 0 \le \lambda \le 1$ and are conveniently 
expressed via an angle $\phi$ which satisfies $\lambda = {\rm cos}\phi$,
\bea
e_1 &=& -{\rm cos}(\phi/3) \nn
e_2 &=& \frac{1}{2}{\rm cos}(\phi/3) - \frac{\sqrt{3}}{2}{\rm sin}(\phi/3) \nn
e_3 &=& \frac{1}{2}{\rm cos}(\phi/3) + \frac{\sqrt{3}}{2}{\rm sin}(\phi/3)
\label{eqD14}
\eea
The solution~(\ref{eqD13}) oscillates between $e_2$ and $e_3$ with a period
equall to $T = 2 K(k)/\alpha$ ($K(k)$ is the complete integral of the first 
kind of elliptic modulus $k$). The angular field $\theta$ is determined from
\be
\theta = \frac{\sqrt{\lambda}}{2} \int \frac{d\xi}{u}
\label{eqD15}
\ee
and using properties of the Jacobi elliptic functions is written as
\be
\theta(\xi) =  \frac{\sqrt{\lambda}}{2 e_1} \Big[\xi -\frac{e_2-e_1}{\alpha e_2}
\Pi(\frac{e_1 k^2}{e_2} ; {\rm am}(\alpha \xi;k^2); k^2) \Big].
\label{eqD16}
\ee
$\Pi (n;x;k^2)$ denotes the incomplete elliptic integral of the third kind
with modulus $k$ and characteristic $n$ while ${\rm am}(x;k^2)$ is the 
Jacobi amplitude function, which satisfies ${\rm sin}({\rm am}(x;k^2))  = {\rm sn}(x;k^2)$. 
A periodic solution for the gauge fields is equivalent to
a {\it closed orbit} for the rotor~(\ref{eqD9}) on the $(r,\theta)$ plane. 
Thus a ``quantization'' condition on the parameter $\lambda$ of the solution
is enforced from the periodicity of $\theta$ for integers $N, N'$ such that:
\be
\theta(N T) = 2 \pi N' \;.
\label{eqD17}
\ee
This, in turn, leads to the highly non-linear relation     
\be
N \frac{\sqrt{\lambda}}{e_1 \alpha} \Big[K(k) -\frac{e_2-e_1}{e_2}
\bar{\Pi}(\frac{e_1 k^2}{e_2} ; k^2) \Big]= 2 \pi N'\;.
\label{eqD18}
\ee
$\bar{\Pi} (n;k^2)=\Pi (n;\pi/2;k^2)$ here denotes the complete elliptic integral of the third kind.
Therefore we look for all values of the angle $\phi$ in $(0, \pi/2)$ such that the function
\be
{\cal{Q}}(\phi) = \frac{\sqrt{\lambda}}{2 \pi e_1 \alpha} \Big[K(k) -\frac{e_2-e_1}{e_2}
\bar{\Pi}(\frac{e_1 k^2}{e_2} ; k^2) \Big] 
\label{eqD19}
\ee
takes the value $N'/N$, i.e. on the set of rationals. From known properties and a Taylor analysis of ${\cal{Q}}(\phi)$ for $\phi$ near $\pi/2$ we deduce that ${\cal{Q}}(\phi)$ increases
monotonically in the $(1/\sqrt{6}, 1/2)$ interval (see Fig.~\ref{fig:fig1}).
Effectively, Eq.(\ref{eqD18}) becomes  a ``quantization'' condition on the angular momentum $L$ of the rotor since $\lambda= 4 L^2$. Solutions of~(\ref{eqD18}) are obtained easily numerically by selecting pairs of integers $(N',N)$ such that $1/\sqrt{6} <  N'/N < 1/2$. The lowest pairs of integers satisfying Eq.(\ref{eqD18}) are shown in Table~\ref{table1}. For a given (large) $N$, one expects
based on the density of primes that the number of solutions are 
roughly $\sim 0.1 N/ \log N$.
\begin{table}
\vspace{0.5cm}
\begin{tabular}{ccccc}
$N'$ & $N$ & $\phi \,({\rm deg}) $ & $k^2$ & $L$ \\
\hline \hline
3 & 7 & 61.0208 & 0.352594 & 0.348027 \\
\hline
4 & 9 & 74.9299 & 0.423877 & 0.254951 \\
\hline
5 & 11 & 80.5348 & 0.452263 & 0.202761 \\
\hline
5 & 12 & 41.9829 & 0.251578 & 0.431087 \\
\hline
6 & 13 & 83.4493 & 0.466980 & 0.168880 \\
\hline
\end{tabular}
\caption{Solutions of Eq.~(\ref{eqD18}) obtained numerically}.
\label{table1}
\end{table}

\nin
The elliptic modulus $k$ as determined from eq.~(\ref{eqD13}) takes an enumerable, infinite set of 
values in the interval $ 0 \le k^2 \le 1/2$. 
A particularly interesting solution is represented by the circular orbit,
$\lambda =1, (\phi=0)$, which has the maximal angular momentum, $L=1/2$.  
At this point $k=0$ and the Jacobi elliptic functions become harmonic
$({\rm sn}(\xi;0) ={\rm sin}\;\xi, {\rm cn}(\xi;0) ={\rm cos}\;\xi,
  {\rm dn}(\xi;0) = 1 )$. 
Since $r = 1/\sqrt{2}$ and $\theta = \xi$, a massive, harmonic wave solution
of the interacting YM EoM is given on the proper frame by~\footnote{We include also arbitrary phase shifts $\phi_1, \phi_2, \phi_3$ on the complex pairs.}
\be
\vec{{\bf A}} = \frac{1}{\sqrt{2}}\begin{pmatrix}
0         & e^{-i mt- i \phi_1}\hat{e}_1   & e^{i m t + i \phi_2}  \hat{e}_2 \\ 
e^{i mt+ i \phi_1} \hat{e}_1    & 0          & e^{-i m t- i \phi_3} \hat{e}_3 \\
e^{-i mt-i \phi_2 } \hat{e}_2  &e^{i mt + i \phi_3} \hat{e}_3      & 0 
\end{pmatrix}
\label{eqD20}   
\ee
The other limiting solution is obtained for $\phi = \pi/2$, 
$(\lambda = 0, L = 0$) where the elliptic modulus takes the value 
$k=1/\sqrt{2}$.  In this limit the solution degenerates to a straight line on 
the plane ($\theta ={\rm const}$). From known properties it can be shown that
\be
r(\xi) = \Big(\frac{\sqrt{3}}{2}\Big)^{1/2} {\rm cn}(3^{1/4}\xi + K(1/\sqrt{2}) ;1/2) \;.
\ee

\begin{figure}[htbp]
\centerline{\includegraphics[width=\columnwidth]{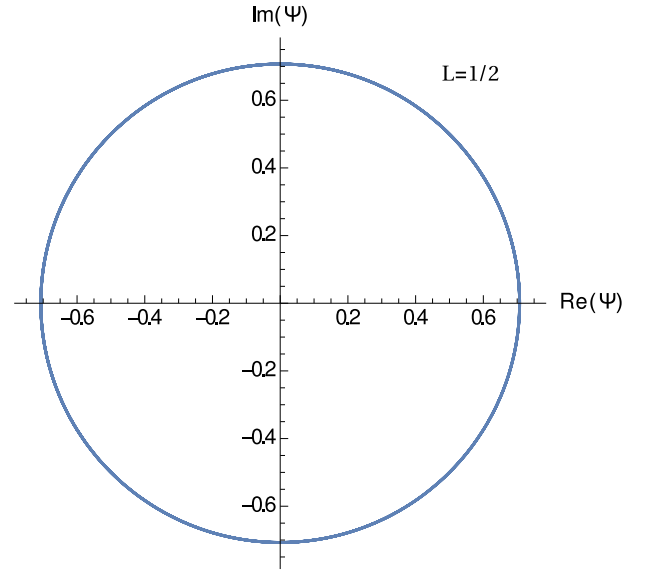}}
\caption{The harmonic plane-wave solution for $L=1/2$ corresponds to the circular orbit on the plane.}
\label{fig:fig2}
\end{figure}

\begin{figure}[htbp]
\centerline{\includegraphics[width=\columnwidth]{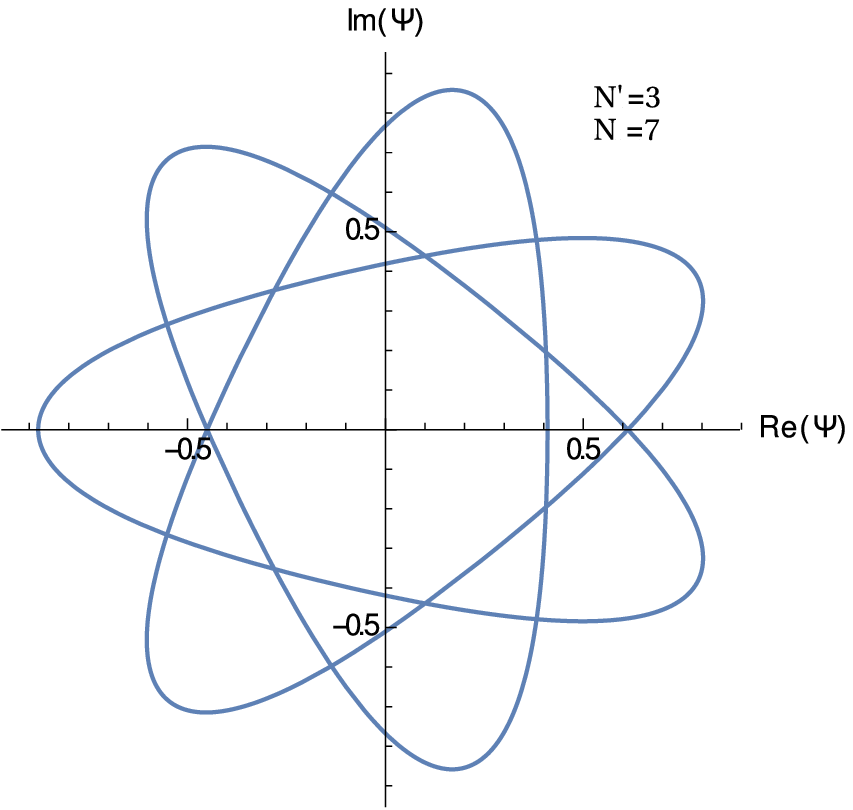}}
\caption{The non-linear plane wave solution, Eq.(\ref{eqD22}), with $N'=3, N=7$. }
\label{fig:fig3}
\end{figure}

\begin{figure}[htbp]
\centerline{\includegraphics[width=\columnwidth]{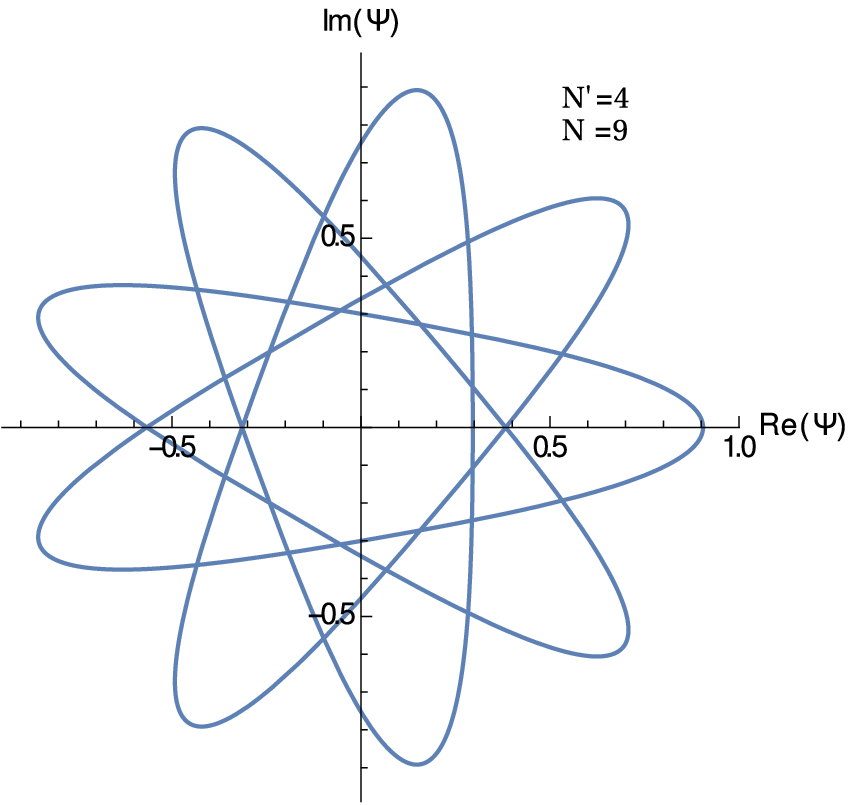}}
\caption{The non-linear plane wave solution, Eq.(\ref{eqD22}), with $N'=4, N=9$. }
\label{fig:fig4}
\end{figure}

\begin{figure}[htbp]
\centerline{\includegraphics[width=\columnwidth]{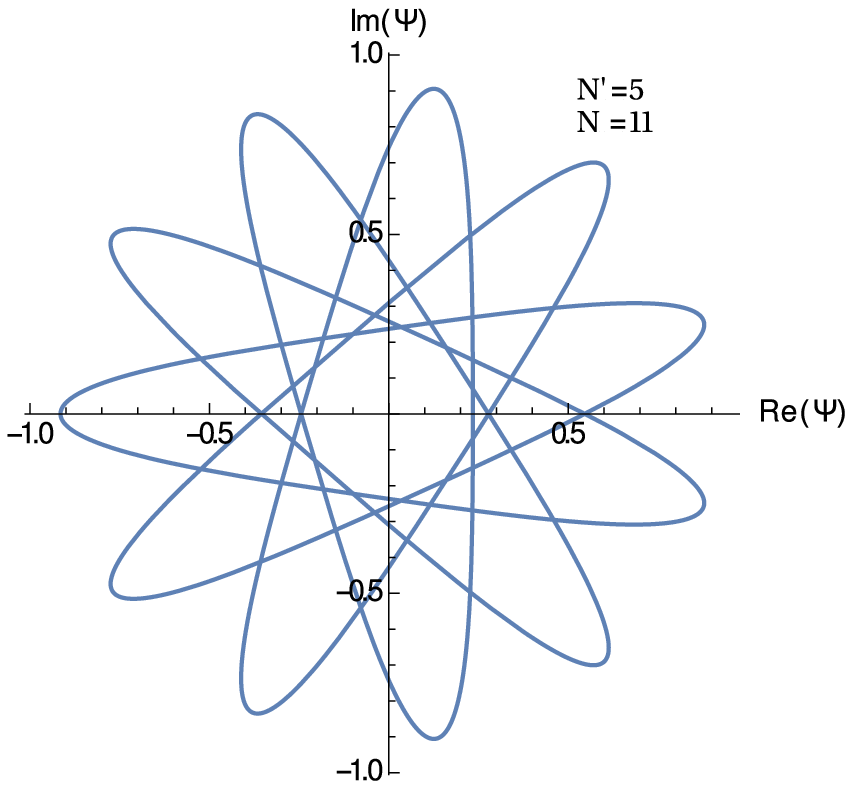}}
\caption{The non-linear plane wave solution, Eq.(\ref{eqD22}), with $N'=5, N=11$. }
\label{fig:fig5}
\end{figure}

On a general frame, the solution is obtained by a Lorentz boost, eqs~(\ref{eqA10}-\ref{eqA11}). The $R^3$ basis ($\hat{e}_1, \hat{e}_2, \hat{e}_3$) is boosted
to the three orthonormal spacelike~\footnote{ Note that 
$\varepsilon^{(\sigma)}_\mu \varepsilon^{\mu (\sigma') } = -\delta^{\sigma \sigma'}$ and $ \varepsilon^{(\sigma)}_\mu \Bbbk^{\mu}=0 $ hold.}
polarization vectors $\varepsilon^{(\sigma)}_\mu$ given for $\sigma=1,2,3$ and $\mu = 0,1,2,3$ 
 \be
\varepsilon^{(\sigma)}_\mu=\begin{pmatrix}
-k_1/m & -k_2/m & -k_3/m \\ 1 + \frac{k_1 ^2}{m(\omega + m )} & \frac{k_1 k_2}{m(\omega + m) } & \frac{k_1  k_3}{m(\omega + m) } \\ \frac{k_1  k_2}{m(\omega + m) } &  1 + \frac{k_2 ^2}{m(\omega + m) } & \frac{k_2  k_3}{m(\omega + m )} \\ \frac{k_1  k_3}{m(\omega +  m )} & \frac{k_2  k_3}{m(\omega + m) } & 1 + \frac{k_3 ^2}{m(\omega + m) }
\end{pmatrix}
\label{eqD21}   
\ee
and the color fields are written
\bea
A^1_\mu + i A^2_\mu &=&2 \varepsilon^{(1)}_\mu r(\xi) e^{i \theta(\xi)+ i \phi_1} \nn
A^4_\mu - i A^5_\mu &=&2 \varepsilon^{(2)}_\mu r(\xi) e^{i \theta(\xi)+ i \phi_2} \nn
A^6_\mu + i A^7_\mu &=&2 \varepsilon^{(3)}_\mu r(\xi) e^{i \theta(\xi)+ i \phi_3} 
\label{eqD22}
\eea
for any selected $L$ such that eq.~(\ref{eqD18}) is satisfied. The harmonic plane 
wave solution is recovered for $r=1/\sqrt{2}$ and $\theta = \omega t   
- \vec{k} \cdot \vec{x} $. Note that the solution~(\ref{eqD22}) satisfies the Lorentz gauge condition $\dmu A^{\mu a} = 0$ {\it on any frame} and this is a direct
consequence of the ansatz~(\ref{eqC1}) chosen on the proper frame. It may also be called 'diagonal' in the sense that the complex $SU(3)$ algebra roots are aligned with the three gluon polarization states.  
Global $SU(3)$ transformations on the solution~(\ref{eqD22}) are allowed since they do not spoil the Lorentz gauge condition. They rotate the solutions
\be
{\bf A}_\mu \rightarrow g {\bf A}_\mu g^{-1}
\ee
--where $g$ is a constant $SU(3)$ matrix in the fundamental-- and thus generate additional color fields in the Cartan subalgebra $(T^3, T^8)$.
The general field component $A_\mu^a$ is a linear superposition of the 'diagonal' solution (\ref{eqD22}) with weights equal to the adjoint matrix elements
$R^{ab}(g)$.

\subsection{$L = 0$ solutions}

\nin
Solutions satisfying $L = 0$ are also possible. In that case the angles $\theta_1, \theta_2, \theta_3$ are necessarily constants. 
\begin{itemize}
\item{The case $\Psi_1 = \Psi_2 = \Psi_3 = 0$ does not lead to interesting 
(periodic) solutions.}
\item{The case $\Psi_1 \ne 0 , \Psi_2 = \Psi_3 = 0$, from the 'Group 2' equations, in order to avoid unbounded solutions for $D_2$ and $D_3$ leads also to
$D_1=D_2=D_3=0$, and finally forbids any bounded solution.} 
\item{The case $\Psi_1 \ne 0 , \Psi_2 \ne 0, \Psi_3 = 0$, from the 'Group 3' equations, leads to $D_1=D_2=0$. This allows the following set of coupled equations for
the remaining fields:
\bea
\ddot{r_1} + r_1 \big[ r_2^2 + D_3^2 \big] &=& 0 \nn
\ddot{r_2} + r_2 \big[ r_1^2 + D_3^2 \big] &=& 0 \nn
\ddot{D_3} + 6 D_3 r_1^2  &=& 0 \nn
\ddot{D_3} + 6 D_3 r_2^2  &=& 0 
\label{eqE1}
\eea
In general this is a chaotic system but the following integrable cases are
included:
\bea 
&&r_1 = r_2 = {\rm cn}(\xi;\frac{1}{2}) \,,\, D_3 = 0 \nn
&&r_1 = r_2 = \frac{D_3}{\sqrt{5}} = {\rm cn}(\sqrt{6}\xi;\frac{1}{2}) 
\label{eqE2}
\eea
}
\item{The case $\Psi_1 \ne 0 , \Psi_2 \ne 0, \Psi_3 \ne 0$ leads to the coupled system
\bea
\ddot{r_1} + r_1 \big[ r_2^2 + r_3^2 \big] &=& 0 \nn
\ddot{r_2} + r_2 \big[ r_1^2 + r_3^2 \big] &=& 0 \nn
\ddot{r_3} + r_3 \big[ r_1^2 + r_2^2 \big] &=& 0 
\label{eqE3}
\eea
which in general presents chaotic behavior. Integrable cases are 
\bea 
&&r_1 = r_2 = r_3 = {\rm cn}(\sqrt{2}\xi;\frac{1}{2}) \nn
&&r_1 = r_2 = {\rm cn}(\xi;\frac{1}{2}) \;, \; r_3 = 0
\label{eqE4}
\eea
and the similar permutations.  
}
\end{itemize}
From the above analysis we conclude that the $L=0$ solutions are less interesting and are always of the ${\rm cn}(\xi;1/2)$ -type which solves also the $SU(2)$ theory.

\section{Other Gauge Groups}

\subsection{SU(2)}

\nin
For $SU(2)$ we solve the Gauss' law via the following ansatz:
\be
\vec{{\bf A}} = \frac{1}{2}\begin{pmatrix}
D \hat{e}_3         & \Psi_{1}^* \hat{e}_1 + \Psi_{2}  \hat{e}_2 \\ 
\Psi_{1} \hat{e}_1  + \Psi_{2}^* \hat{e}_2      & -D \hat{e}_3 
\end{pmatrix}
\label{eqF1}   
\ee
with $D$ a real function and $\Psi_1, \Psi_2 $ complex functions of $\xi$.
The Gauss law is written
\be
G^a T^a = \vec{{\bf A}}\cdot \dot{\vec{{\bf A}}} - \dot{\vec{{\bf A}}} \cdot \vec{{\bf A}} =
\frac{i}{2} \begin{pmatrix}
L_1-L_2 & 0 \\
0 & L_2 - L_1 \end{pmatrix}
\label{eqF2}   
\ee
where we introduced
\be
L_j = \frac{i}{2} (\Psi_j \dot{\Psi_j^*} - \Psi_j^* \dot{\Psi_j}) \;, j = 1,2. 
\label{eqF3}
\ee 
Implementation of the Gauss law requires $ L_1 = L_2 = L$.
The EoM for the ansatz~(\ref{eqF2}) are:
\bea
\ddot{D}+ D \Big[ |\Psi_1|^2 + |\Psi_2|^2 \Big] = 0 \nn
\ddot{\Psi_1}+\Psi_1 D^2 + \frac{1}{2}\Psi_2^* \Big[\Psi_1 \Psi_2 - \Psi_1^* \Psi_2^*\Big] = 0 \nn
\ddot{\Psi_2}+\Psi_2 D^2 + \frac{1}{2}\Psi_1^* \Big[\Psi_1 \Psi_2 - \Psi_1^* \Psi_2^*\Big] = 0 
\label{eqF4}   
\eea
The above equations do not admit harmonic plane-wave solutions. Assuming the ansatz
$\Psi_1 = c_1 e^{i \omega_1 \xi}, \Psi_2 = c_2 e^{i \omega_2 \xi} $ with
constants $c_1,c_2, \omega_1, \omega_2$ and enforcing the Gauss' law
$ L_1 = L_2 $ it is shown easily that such solutions are not allowed. The only
integrable case of~(\ref{eqF4}) appears for $ \Psi_1 = i \Psi_2 = D$ which is non-other than the original solution in~\cite{Savvidy1}.

\subsection{SU(4)}
\nin
For the case of $SU(4)$ gauge theory we propose the following ansatz for the gauge potential which similar to eq.~(\ref{eqC1}) has by construction 
orthogonal rows (columns) to each other:
\be
\vec{{\bf A}} = \begin{pmatrix}
0  & \Psi_{1} \hat{e}_1  & \Psi_{2}  \hat{e}_2  & \Psi_{3} \hat{e}_3 \\    
\Psi_{1}^* \hat{e}_1  & 0  & \Psi_{6}  \hat{e}_3  & \Psi_{5} \hat{e}_2 \\    
\Psi_{2}^* \hat{e}_2  & \Psi_{6}^* \hat{e}_3    & 0 & \Psi_{4} \hat{e}_1 \\    
\Psi_{3}^* \hat{e}_3  & \Psi_{5}^*  \hat{e}_2  & \Psi_{4} \hat{e}_1 & 0
\end{pmatrix}
\label{eqG1}   
\ee
The Gauss law for the above ansatz is written
\bea
&&G^a T^a =\vec{{\bf A}}\cdot \dot{\vec{{\bf A}}} - \dot{\vec{{\bf A}}} \cdot \vec{{\bf A}} 
=-2 i \;{\rm diag} 
\Big(L_1+L_2+L_3 , \nn
&&-L_1+L_5+L_6 ,-L_2+L_4-L_6,-L_3+L_4-L_5 \Big)  
\label{eqG22}   
\eea
where we introduced 
\be
L_j = \frac{i}{2} (\Psi_j \dot{\Psi_j^*} - \Psi_j^* \dot{\Psi_j})  \;, j= 1,2,...,6. 
\label{eqG3}
\ee 
Implementation of the Gauss law requires:
\bea
L_1+L_2+L_3 = 0   &&, \;\;\;\; -L_1+L_5+L_6 = 0  \nn
-L_2+L_4-L_6 = 0   &&, \;\;\;\;-L_3+L_4-L_5 = 0
\label{eqG4}   
\eea
The EoM contain cubic terms of the fields $\Psi_i$
and a general solution goes beyond the scope of this work. 
Here, we simply note that $SU(3)$ solutions can be readily 
immersed into $SU(4)$ in four different ways:
\bea
\Psi_1=\Psi_2=\Psi_3=0 \;, \;\;\Psi_1=\Psi_5=\Psi_6=0 \nn
\Psi_2=\Psi_4=\Psi_6=0 \;, \;\;\Psi_3=\Psi_4=\Psi_5=0 
\eea
For each of the choises, the remaining three fields are identified with the
$SU(3)$ solution, Eq.(\ref{eqD22}). Thus, plane waves, and particularly 
harmonic,
can also propagate in the $SU(4)$ theory, limited --at least-- in the $SU(3)$ subspaces.

\nin

\section{Conclusions and Outlook}

We have presented plane wave solutions for the $SU(3)$ Yang-Mills EoM in $3+1$ dimensions. The solutions represent massive non-linear plane waves, for arbitrary values of the mass parameter --reflecting the scale invariance of the action-- and obeying the relativistic dispersion relation. The solution is designed by alignining the non-diagonal color fields with the polarization states of the massive vector boson. Via a global 
$SU(3)$ transformation all the octet fields participate in the solution.
The functional form is described by the dynamics of a planar particle bounded with the $r^4$ potential. The periodic particle orbits on the plane are characterized from the value of the angular momentum $L$ which is bounded (in scaled units) between zero and $1/2$. The value of angular momentum is related to the elliptic modulus $k$ of the Jacobi elliptic functions which describe the solution. Each rational number in the $(1/\sqrt{6},1/2)$ interval
gives a solution, and at the edge of the interval, with $L=1/2$,  a harmonic massive plane-wave solution of the {\it interacting} YM theory is recovered. Compared to the $SU(2)$ theory which possesses only the $k^2=1/2$ plane wave, the $SU(3)$ theory presents a far more rich spectrum of solutions with $k^2$ obtaining an infinite, enumerable set of values covering densely the interval $0 \le k^2 \le 1/2$.

The coupling to fermions is straightforward for static quark matter in the Cartan  $(T^3,T^8)$ subalgebra. The Gauss law, Eq.~(\ref{eqC3}), admits quark densities on the {\it r.h.s} if angular momenta values $L_1 \ne L_2 \ne L_3$ are used.

Plane wave solutions, and in particular the harmonic ones, may be of use in quantization schemes or particular perturbative treatments of the quantum theory since
they incorporate automatically a gluon mass. The impact of such configurations to the properties of the quantum theory is worthwhile to assess. 
In addition, classical Minkowskian solutions may also be useful in
the study  of gluon radiation, or the thermodynamical properties near the phase transition where semiclassical configurations become relevant. 

Finally, we note that higher rank gauge groups admit similar treatment. For the $SU(4)$ theory, the Gauss law can be solved following the same strategy. In the $SU(3)$ subgroups the plane wave solutions remain valid, while the issue of existence of more generic $SU(4)$ solutions is left for future investigations.


\begin{thebibliography}{99}
\bibitem{books} S. Coleman, {\it Aspects of Symmetry} (Cambridge University Press, Cambridge, England, 1985); R. Rajaraman, {\it Solitons and Instantons} (North-Holland, Amsterdam, 1982).
\bibitem{Coleman1} S. Coleman, Comm. Math. Phys. {\bf 55}, 113 (1977).
\bibitem{Savvidy1} G. Z. Baseyan, S. G. Matinyan and G. K. Savvidy, Pis'ma Zh. Eksp. Teor. Fiz. {\bf 29}, 641 (1979); JETP Lett. {\bf 29}, 588 (1979).
\bibitem{Corrigan} E. Corrigan and D. B. Fairlie, Phys. Lett. {\bf B 67}, 69 (1977).
\bibitem{Ohteh1} C. H. Oh and R. Teh, Phys. Lett. {\bf B 87}, 83 (1979).
\bibitem{Ohteh2} C. H. Oh and R. Teh, J. Math. Phys. {\bf 26}, 841 (1985).
\bibitem{Coleman2} S. Coleman, Phys. Lett. {\bf B 70}, 59 (1977).
\bibitem{melia} F. Melia and S. Lo, Phys. Lett. {\bf B 77}, 71 (1978).
\bibitem{Frasca} M. Frasca, Phys. Lett. {\bf B 670}, 73 (2008); Mod. Phys. Lett. A {\bf 24}, 2425 (2009).
\bibitem{Smilga} A. Smilga, {\it Lectures on Quantum Chromodynamics} (World Scientific, Singapore, 2001).
\bibitem{Savvidy2} S. G. Matinyan, G. K. Savvidy and N. G. Ter-Arutyunyan-Savvidy, Sov. Phys. JETP {\bf 53}, 421 (1981); Sov. Phys. JETP {\bf 34}, 590 (1981).
\bibitem{Abram} M. Abramowitz and I. A. Stegun, {\it Handbook of Mathematical Functions} (Editors) (1964); I. S. Gradshteyn and I. M. Ryzhik, {\it Table of Integrals, Series and Products} (Academic Press, USA, 2007).

\end{thebibliography}
\end{document}